\documentclass[10pt, conference, letterpaper]{IEEEtran}

\usepackage{cite}
\ifCLASSINFOpdf

\else

\fi
\usepackage{subfig}
\usepackage{bm}
\usepackage{amsmath}
\usepackage{epsf}
\usepackage{graphics}
\usepackage{ amssymb }
\usepackage[dvips]{graphicx}
\usepackage{epsfig}
\usepackage{cite}
\usepackage[linesnumbered,ruled,vlined]{algorithm2e}
\usepackage{colortbl}
\usepackage{color}
\usepackage{multicol}
\usepackage{bm}
\usepackage{amsmath}
\usepackage{epsf}
\usepackage{graphics}
\usepackage{ amssymb }
\usepackage[dvips]{graphicx}
\usepackage{epsfig}
\usepackage{cite}
\usepackage[linesnumbered,ruled,vlined]{algorithm2e}
\usepackage{graphicx}
\usepackage{epsfig}
\usepackage{latexsym}
\usepackage{amsfonts}
\usepackage{here}
\usepackage{rawfonts}
\usepackage{subfig}

\usepackage{bm}
\usepackage{amsmath}
\usepackage{epsf}
\usepackage{graphics}
\usepackage{ amssymb }
\usepackage[dvips]{graphicx}
\usepackage{epsfig}
\usepackage{cite}
\usepackage[linesnumbered,ruled,vlined]{algorithm2e}
\usepackage{colortbl}
\usepackage{color}
\usepackage{multicol}
\usepackage{bm}
\usepackage{amsmath}
\usepackage{epsf}
\usepackage{graphics}
\usepackage{ amssymb }
\usepackage[dvips]{graphicx}
\usepackage{epsfig}
\usepackage{cite}
\usepackage[linesnumbered,ruled,vlined]{algorithm2e}
\usepackage{graphicx}
\usepackage{epsfig}
\usepackage{latexsym}
\usepackage{amsfonts}
\usepackage{here}
\usepackage{rawfonts}

\usepackage[utf8]{inputenc}
\usepackage[english]{babel}
\usepackage{amsmath}
\usepackage{amsfonts}
\usepackage{amssymb}
\usepackage{color} 
\usepackage{bm}
\usepackage{listings}

\usepackage{amssymb}
\usepackage{amsthm}
\usepackage{graphicx}
\usepackage{epstopdf}
\usepackage{listings}
\usepackage{float}
\usepackage{amsmath}
\usepackage{amssymb}
\usepackage{amsfonts}
\usepackage{epstopdf}

\usepackage{multirow}
\usepackage{amscd}
\usepackage{mathrsfs}
\usepackage{graphicx}
\usepackage{color}
\usepackage{url}
\usepackage{bm}
\usepackage{setspace}

\usepackage{footnote}
\usepackage[utf8]{inputenc}
\usepackage[english]{babel}
\usepackage{amsmath}
\usepackage{amsfonts}
\usepackage{amssymb}
\usepackage{color} 
\usepackage{bm}
\usepackage{listings}

\usepackage{amssymb}
\usepackage{amsthm}
\usepackage{graphicx}
\usepackage{epstopdf}
\usepackage{listings}
\usepackage{float}
\usepackage{amsmath}
\usepackage{amssymb}
\usepackage{amsfonts}
\usepackage{epstopdf}

\usepackage{multirow}
\usepackage{amscd}
\usepackage{mathrsfs}
\usepackage{graphicx}
\usepackage{color}
\usepackage{url}
\usepackage{bm}
\usepackage{setspace}

\usepackage{footnote}
\usepackage[]{algorithm2e}
\renewcommand{\paragraph}{\@startsection{paragraph}{4}{\z@}%
	{-2.00ex\@plus -1ex \@minus -.2ex}%
	{0.5ex \@plus .2ex}%
	{\normalfont\normalsize}}
	
 \usepackage[protrusion=true,expansion=true]{microtype}

\IEEEoverridecommandlockouts
\usepackage{cite}
\usepackage{amsmath,amssymb,amsfonts}
\usepackage{algorithmic}
\usepackage{graphicx}
\usepackage{textcomp}
\usepackage{xcolor}

\usepackage[]{algorithm2e}
\renewcommand{\paragraph}{\@startsection{paragraph}{4}{\z@}%
	{-2.00ex\@plus -1ex \@minus -.2ex}%
	{0.5ex \@plus .2ex}%
	{\normalfont\normalsize}}
\RequirePackage{fix-cm} 

\makeatletter 
\newcommand\semiHuge{\@setfontsize\semiHuge{22.72}{27.38}}

\begin{document}

\title{{  \semiHuge{ Dynamic  Mobility-Aware Interference Avoidance  for Aerial Base Stations in Cognitive Radio Networks}} \thanks{This work was supported in part by the National Science Foundation under grants ECCS-1444009 and CNS-1824518.} }

\author{\IEEEauthorblockN{Ali Rahmati$^*$, Xiaofan He$^\dagger$, Ismail Guvenc$^*$, and Huaiyu Dai$^*$}
\IEEEauthorblockA{{$^*$Department of Electrical and Computer Engineering}, 
{NC State University,}
Raleigh, NC, US \\ $^\dagger$School of Electronic Information, Wuhan University, Wuhan, China\\
Email:\texttt{ \{arahmat@ncsu.edu, xiaofanhe@whu.edu.cn, iguvenc@ncsu.edu, hdai@ncsu.edu}\}}
}
\maketitle

\begin{abstract}
Aerial base station (ABS) is a promising solution for public safety as it can be deployed in coexistence with cellular networks to form a temporary communication network. However, the interference from the primary cellular network may severely degrade the performance of an ABS network. With this consideration, an adaptive dynamic interference avoidance scheme is proposed in this work for ABSs coexisting with a primary network. In the proposed scheme, the mobile ABSs can reconfigure their locations to mitigate the interference from the primary network, so as to better relay the data from the designated source(s) to  destination(s). To this end, the single/multi-commodity maximum flow problems are formulated and the weighted Cheeger constant is adopted as a criterion to improve the maximum flow of the ABS network. In addition, a distributed algorithm is proposed to compute the optimal ABS moving directions. Moreover, the trade-off between the maximum flow and the shortest  path trajectories is investigated and an energy-efficient approach is developed as well. Simulation results show that  the proposed approach is effective in improving the maximum network flow  and the energy-efficient approach can save up to $39\%$ of the energy  for the ABSs with marginal degradation in the maximum network flow. 
\end{abstract}


\vspace{-3mm}
\section{Introduction}
 {Recently}, aerial base stations (ABSs) have been  introduced as an effective solution for a variety of important applications, including environmental monitoring, public safety, disaster relief, search/rescue,  surveillance, and purchase delivery~\cite{hayat2016survey, gupta2016survey}. The federal aviation administration (FAA) reported that the number of ABSs will exceed 7 million in 2020 \cite{mozaffari2018tutorial}. The ABSs  can be implemented using  unmanned aerial vehicles (UAVs). Being cost-effective, flexible and suitable for low-latency communication in 5G \cite{8367785,tavana2018congestion}, the ABSs are considered as a promising supplement  to  conventional terrestrial base stations to provide broadband connectivity in complex scenarios. 

Nonetheless, forming an ABS communication network for the aforementioned applications is highly non-trivial. Particularly, since there is no regulation and pre-allocated spectrum band for the ABSs \cite{stocker2017review , {saleem2015integration}}, in most  scenarios the ABS network may need to share the spectrum with an existing primary network. Without  careful management, the performance of both the ABS network and the primary network may be considerably degraded due to  mutual interference. This clearly suggests a compelling need for developing innovative ABS positioning and mobility management schemes.

With this consideration, the interference avoidance problem for the ABS network formation is studied in this work, with the idea that the mobility feature of ABSs offers a built-in advantage for dynamic re-positioning and thus interference avoidance.
Specifically, a novel dynamic mobility-aware interference avoidance scheme is proposed in this work by leveraging spectral graph theory and incorporating practical ABS channel models. Through the proposed scheme, the ABSs can dynamically reconfigure their locations to avoid the interference caused by the primary network and optimize the data flow from the designated source(s) to destination(s).


The contributions of this work can be summarized as follows:
\begin{itemize}
\item  ABS network interference avoidance  is
formulated as a single/multi-commodity maximum flow problem.

\item Taking  advantage of the inherent mobility feature of the ABSs and incorporating practical ABS channel model,  an adaptive interference avoidance scheme  is developed. Using the proposed scheme, the ABSs can effectively reconfigure their positions to avoid the interference and improve the data flow from the source(s) to the destination(s). 

\item The weighted Cheeger constant for 3D scenarios is introduced  as an effective metric to improve the ABS network connectivity. 

\item A distributed algorithm for computing  the optimal ABS moving directions is presented.
\item The trade-off between  energy efficiency and the maximum flow of the ABS network is investigated. 
\end{itemize}

The rest of the paper is organized as follows. In Section~\ref{relw}, the related works are discussed. In Section~\ref{sec2}, the system model is presented.
In Section~\ref{sec3}, the single-commodity maximum flow problem is formulated and the weighted Cheeger constant is adapted to 3D coordinates. The multi-commodity problem is presented in Section~\ref{sec4}. In Section~\ref{sec5}, several practical considerations related to ABS deployment  are discussed. Section \ref{sec7} investigates the performance and effectiveness of the proposed scheme via extensive simulations, and finally, Section~\ref{sec8} concludes the paper.

\section{Related Works}\label{relw}
\vspace{-1mm}

\noindent Due to the lack of regulation and  spectrum scarcity, the ABSs may have to coexist with a primary network in practice. Thus, ABS network management  is necessary to mitigate the interference from/to the primary network. In \cite{8292271}, the energy efficiency of a cognitive
radio ABS is investigated in the presence of primary transceivers.  In \cite{athukoralage2016regret}, a dynamic transmission duty cycle allocation scheme  is proposed for coexisting ABS and cellular/WiFi networks based on regret game. In addition, it was shown that the duty cycle allocation strategy based on the correlated equilibrium of the game outperforms  static duty cycle allocation. In spite of these pioneering works,    the dynamic ABS positioning problem in the cognitive radio setting has not been considered, where the key research problem is  how to design the trajectories of the ABSs  to mitigate the mutual interference between the ABS network and the primary network.

 The ideas of mobility-aware interference-avoidance, path planning and trajectory optimization  have already been explored for ABS networks in the literature \cite{zeng2017energy, mozaffari2018beyond, rohde2012interference, challita2018artificial,abdel2017optimum,han2009optimization}. However,  mobility-aware interference-management becomes more challenging in the cognitive radio setting and so is the design of effective interference avoidance trajectories for the ABS.
 The authors in \cite{zeng2017energy} proposed  energy-efficient  ABS trajectory optimization for given ABS initial and final locations, and showed that this   scheme  outperforms   other existing schemes, including those based on
 heuristic rate-maximization and energy-minimization, in terms of ABS communications. In contrast, the final locations of the ABSs  in our work are assumed unfixed and derived towards the maximum flow metric. In \cite{rohde2012interference}, the effectiveness of ABS deployment in overload and outage situations is investigated where the
relay placement, the number of relays, and the relay transmit
power are analyzed. In \cite{abdel2017optimum,han2009optimization, ig1,ig2}, the connectivity of the ABS network is  
characterized by its Fiedler value and algebraic connectivity. However, their approach is different from ours, as they consider neither the cognitive radio setting nor dynamic interference avoidance.


In \cite{7330770}, the importance of algebraic connectivity for controlling the positions of ABSs towards a desired
formation is shown. A mixed-integer linear
programming is formulated in order to maximize the algebraic connectivity
 and its upper bounds are obtained based on
cutting plane methods.  In  \cite{6315537}, ABS-based monitoring applications are considered,
where  ABSs are deployed to relay \cite{rahmati2015price, wu2018joint, javan2017resource} and
transmit time  information between  the vehicles
and the control stations.
 The key difference between  \cite{7330770,6315537}   and our work is that we  consider mobile ABSs to dynamically improve the network connectivity  in a cognitive radio setting.

\section{System Model}\label{sec2}
\subsection{Scenario Description}
\noindent Consider a scenario in which a set of sources of interference~(SIs) (e.g., base stations, small cells, or user equipment in the primary network \cite{demestichas20135g,khamidehi2016joint}) and ABSs coexist in the network. The ABSs aim to form a temporary network to relay  data from a designated terrestrial source to a terrestrial destination. To mitigate the interference  from the SIs, the ABS nodes
can change  their geographic locations in the 3D  space. We assume that the SIs can be detected using existing sensing methods as in \cite{tavana2017cooperative}.
 
Assume that there is a set $\mathcal{N}$ of $N$ nodes, including the ABSs, the terrestrial source ($s$), and the destination ($d$)  in the system, in addition to  a set $\mathcal{M}$ of $M$ SIs. The positions of the ABSs and the SIs are denoted by: $\{{\boldmath{r}_i}=(x_i,y_i,h_i) \in \mathbb{R}^3, ~i=1, \dots , N$\} and $\{{\boldmath{r}_p}=(x_p,y_p) \in \mathbb{R}^2, ~ p=1, \dots , M \}$, respectively. Note that  the third coordinates of the terrestrial source and destination are equal to zero. 
In the considered model, the ABS network consists of the source, the destination, and the ABSs.

 Matrix $\mathbf{A}=[a_{i,j}]_{i,j=1}^{N}$ is the so-called generalized adjacency matrix of the ABS network~\cite{chung1997spectral}, in which the elements $a_{i,j}$'s are defined based on the 1-nat information exchanging time between  nodes $i$ and $j$ as:
\begin{equation}
a_{i,j}=\left\{\begin{matrix}
B \left(\frac{1}{\ln(1+\textrm{SIR}_{i,j})}+\frac{1}{\ln(1+\textrm{SIR}_{j,i})} \right)^{-1},& i \ne j\\ 
0\hspace{46mm}, & i=j
\end{matrix}\right.,
\end{equation}
where $B$ is the  total bandwidth of the system, and $\textrm{SIR}_{i,j}$ is the signal-to-interference
ratio at node $j$ for the transmission from node $i$ assuming unit transmission power from the transmitter and interferers. In this work,  $\textrm{SIR}_{i,j}$ is given by:
\begin{equation}
\textrm{SIR}_{i,j}=\frac{ g_{i,j}}{ \sum_{p=1}^{M} g_{p,j}+\sum_{k \in {\mathcal{Q}}_{i,j}}u(d_{j,k}/r_{\textrm{int}})},
\end{equation}
where $g_{i,j}$ stands for the average channel gain between node $i$ and node $j$,  further discussed below; $\mathcal{Q}_{i,j}$ denotes the set of all the nodes excluding node $i$ and $j$, and $d_{i,j}$ is the Euclidean distance between  node $i$ and $j$.
The smoothed step function is defined as: \[u(y)=\zeta \frac{\textrm{exp}(- \kappa y - \log y_0)}{1+\textrm{exp}(- \kappa y - \log y_0)},\]
where $y_0$ is an arbitrarily small positive number
and $\zeta$ and $\kappa$ are design parameters. These parameters together with $r_{\textrm{int}}$ determine the required safety separation of the ABSs in practice. The first term of the denominator in (2) is related to the interference caused by the SIs, while the second term  is included to prevent the ABSs from getting too close to and  colliding with each others. 


Let us define the matrix $\mathbf{D}=\textrm{diag}\{\beta_1, ..., \beta_N\}$ as the generalized degree matrix of the ABS network with $\beta_i= \sum_{\{j|j \ne i\}}a_{i,j}$. Thus, the Laplacian matrix of the network graph is defined as $\mathbf{L}=\mathbf{D}-\mathbf{A}$.
Since  both terrestrial and aerial nodes  are involved, it is necessary to capture the corresponding channel models. In the following, the air to air (A2A) and air  to ground (A2G)   channel models for ABS nodes are presented in details.

\subsection{ABS A2A and A2G Channel Models}
\noindent Denote the set of the aerial nodes by $\mathcal{A}= \mathcal{N}\backslash \{s, d\}$, and the set of terrestrial nodes by $\mathcal{G}=\mathcal{M} \cup \{s, d\}$.
 The wireless links between    node $i \in \mathcal{A}$ (i.e., ABSs)  and  node  $j \in \mathcal{G}$ (i.e., the source, the destination, and the interference sources) can either be line-of-sight (LoS) or non-line-of-sight (NLoS). However, only LoS channels are considered for the links  between the nodes   $i, j \in \mathcal{A}$~\cite{8254715}. In our model, we assume channel reciprocity  for all links.

The signal propagation in A2G channel is mainly affected by the
obstacles and buildings in the environment. Specifically, an A2G communication link
can be either LoS or NLoS with a certain
probability depending on the
location, the height,  the number of obstacles, as well as the
elevation angle between the ABS and the ground
node. In our scheme, we assume a widely-used probabilistic
path-loss model provided by the International Telecommunication
Union (ITU). In particular, the path-loss between
an ABS  node $i \in \mathcal{A}$ and  a ground node $j \in \mathcal{G}$ is given by~\cite{al2014modeling, mozaffari2017mobile}:
\begin{equation}
\Gamma_{i,j}=\left\{\begin{matrix}
(K_o d_{i,j})^\alpha\mu_{\textrm{LoS}},& \textrm{if LoS link} \\ 
~(K_o d_{i,j})^\alpha\mu_{\textrm{NLoS}}, & \textrm{~~if NLoS link},
\end{matrix}\right.
\end{equation}
where
$\mu_{\textrm{LoS}}$ and
$\mu_{\textrm{NLoS}}$
are  different attenuation factors  for the LoS and the NLoS links, respectively,  $f_c$ is the carrier frequency, $\alpha$ is the path-loss exponent, and $c$ is
the speed of light, and   $K_o =  \frac{4\pi f_c}{c}$. 
For  aerial nodes $i \in \mathcal{A}$ and the ground node $j \in \mathcal{G}$, the probability of having an LoS link is given by~{\cite{al2014modeling, mozaffari2017mobile}}:
\begin{equation}
P_{i,j}^{\textrm{LoS}}= \frac{1}{1+\psi \exp(-\eta[\theta_{ij}-\psi])},
\end{equation}
where $\psi$ and $\eta$ are constants  depending the carrier frequency and  the conditions of the environment. The  elevation angle
between an ABS node $i \in \mathcal{A}$ and a ground node $j \in \mathcal{G}$ is given by:
\[\theta_{i,j}=\frac{180}{\pi} \times \sin^{-1}\left(\frac{ \Delta h_{i,j}}{d_{i,j}}\right),\]
where $\Delta h_{i,j}$ is the height difference between  node $i$ and node $j$. Clearly, the probability of NLoS is $P^{\textrm{NLoS}}_{i,j}=1-P^{\textrm{LoS}}_{i,j}.$ The average path-loss of the A2G link from node $i \in \mathcal{A}$ to node $j \in \mathcal{G}$ can be obtained as {\cite{al2014modeling}}:
\[\bar L_{i,j}^{\textrm{A2G}}=(K_o d_{i,j})^\alpha [P^{\textrm{LoS}}_{i,j} \times \mu_{\textrm{LoS}}+P^{\textrm{NLoS}}_{i,j} \times \mu_{\textrm{NLoS}}].\]

As we assume LoS links between different
ABSs, the path-loss model
between ABS $i$ and ABS $j$ is given by {\cite{al2014modeling}}:
\[ L_{i,j}^{\textrm{A2A}}=(K_o d_{i,j})^\alpha \mu_{\textrm{LoS}}.\]
Based on this, the channel power gain between an ABS $i$ and a ground/aerial node $j$  is given by:
\begin{equation}
g_{ij}=\left\{\begin{matrix}
\frac{1}{L_{i,j}^{\textrm{A2A}}},& \textrm{A2A link}, \\ 
\frac{1}{\bar L_{ij}^{\textrm{A2G}}}, & \textrm{A2G link}.
\end{matrix}\right.
\end{equation} 
With the path-loss models for the A2A and A2G links in the system and the network topology, one can obtain the generalized Laplacian matrix of the network.   In the next section, we formulate the single-commodity maximum flow problem for  the considered scenario.

\section{single-commodity Maximum Flow Problem for ABS Network Formation }\label{sec3}

\noindent In this section, the ABS network formation problem for the case of a single pair of source and destination is formulated as a single-commodity maximum flow problem. In this problem, the flow through each edge is upper bounded by the capacity of the corresponding wireless link.
The incoming and outgoing flows for each ABS are assumed equal (i.e., balanced flows).
\subsection{The Single-commodity Maximum Flow Problem}
\noindent Define a directed flow graph 
$G=(\mathcal{N},\mathcal{E})$ where each edge has the capacity $a_{i,j}$, and   $\mathcal{E}$ denotes the set of available edges  in the network. In the single-commodity maximum flow problem, the task is to determine the maximum amount of  flow that can be sent from the source to the destination.

For each link $(i,j)\in \mathcal{E}$, let $f_{i,j}$ denote the flow on link $(i,j)$, which admits $0 \le f_{i,j} \le a_{i,j}$.  Mathematically, the single-commodity maximum flow problem can be defined as:
\begin{equation}\nonumber
\begin{aligned}\label{eqeq}
&~~~ {\text{max}}
& & ~~~~\underset{{j:(s,j) \in \mathcal{E}}}{\sum}{  f_{s,j}}\\
& ~~~~\text{s.t}
& & \underset{\{i:(i,j) \in \mathcal{E}\}}{\sum}{  f_{i,j}}-\underset{\{l:(j,l) \in \mathcal{E}\}}{\sum}{  f_{j,l}} =0, ~~\forall  j \in \mathcal{N}\backslash \{s, d\}, \\& & & 0 \le f_{i,j} \le a_{i,j}, ~ \forall (i,j) \in \mathcal{E}.
\end{aligned}
\end{equation}
The first constraint corresponds to the assumption of balanced flows for all the nodes, except for the source and the destination.
The well-known Ford-Fulkerson algorithm \cite{ford1956maximal} can be employed to solve the above maximum flow problem. 

 
  
 \subsection{Cheeger Constant}

\noindent   For the normalized Laplacian matrix $\mathcal{L}=~\mathbf{D}^{-1/2}\mathbf{L} \mathbf{D}^{-1/2}$, the Cheeger constant is defined as~\cite{chung1997spectral}:
\begin{equation}\label{cheegi}
h(\mathcal{L})=\underset{S}{\text{min}} \frac {\sum_{i \in S, j \in \bar S}a_{i,j}}{\text{min}\{ \textrm{vol}(S),\textrm{vol}(\bar S)\}},
\end{equation}
where $S \subset \mathcal{N}$ is a subset of the nodes and $\textrm{vol}(S)=\sum_{i\in S} \beta_i$.
The well-known normalized Cheeger's inequality is given by~\cite{de2007old}:
\begin{equation}\label{algeb}
\lambda_2(\mathcal{L})/2 \le h(\mathcal{L}) \le \sqrt{2 \lambda_2(\mathcal{L})}.
\end{equation}
In \eqref{algeb},  the second smallest eigenvalue of the network Laplacian matrix $\lambda_2(\mathcal{L})$ is a measure of the connectivity of the graph,  defined as \cite{de2007old}:
\begin{equation}\label{alg121}
\lambda_2(\mathcal{L})= \underset{v\ne \bm{0}, v \perp \bm{1}}{\text{min}} \frac{\langle \mathcal{L}v, v\rangle}{\langle v, v\rangle}, 
\end{equation}
where $\bm{0}$ ($\bm{1}$) is the vector with all coordinates 
equal to $0$ ($1$).
Based on \eqref{algeb}, in order to maximize the Cheeger constant \eqref{cheegi}, one can  maximize the second smallest eigenvalue of the network (also known as the algebraic connectivity).
However, one drawback of $h(\mathcal{L})$  is that it is
invariant to the scaling of the link capacities $a_{i,j}$, while the maximum flow of the network for a given  source-destination pair is obviously not. To address this, we utilize the weighted version of the Cheeger constant~{\cite{6807812}}.
\vspace{-1mm}

\subsection{Weighted Cheeger Constant}\label{cheeg}
\noindent Besides the scaling issue, the regular Cheeger constant  also blindly aims at improving the weakest links in the network and may fail to emphasize the flow for a given source-destination pair. Thus, we need to  distinguish the source and destination nodes from the ABSs nodes, and weighted Cheeger constant comes as a remedy,  defined  as~{\cite{6807812}}:
\begin{equation}
h_{\mathbf{W}}(\mathcal{L})=\underset{S}{\text{min}} \frac {\sum_{i \in S, j \in \bar S}a_{i,j}}{\text{min}\{ |S|_{\mathbf{W}}),(|\bar S|_{\mathbf{W}})\}},
\end{equation}
where the weighted cardinality  $|S|_{\mathbf{W}}=\sum_{i \in S}w_i$, and $w_i~\ge~0$ is the weight of  the node $i$,  adopted to stress the bottleneck for the flow between the source and destination nodes.
The weighted Laplacian matrix is defined as:
\begin{equation}
\mathcal{L}_\mathbf{W}=\mathbf{W}^{-1/2}\mathcal{L} \mathbf{W}^{-1/2}, \textrm{ with}~ \mathbf{W}=\textrm{diag}\{w_1, ...,w_n\}.
\end{equation}
To achieve our goal, the ideal choice is to consider the weights for the source and destination as 1, and those for the ABSs as 0. However, this makes the problem intractable because the weight matrix $\mathbf{W}$ will be non-invertible. As a result, a practical setting is to choose the weights large for the source(s) and destination(s)  and  small enough for the ABSs.
Similar to \eqref{alg121}, the weighted second smallest eigenvalue  $\lambda_2({\mathcal{L}_\mathbf{W}})$ can be defined as:
\begin{equation}
\lambda_2(\mathcal{L}_\mathbf{W})= \underset{v\ne \bm{0}, v \perp \mathbf{W}^{1/2}\bm{1}}{\text{min}} \frac{\langle \mathcal{L}_\mathbf{W} v, v\rangle}{\langle v, v\rangle}. 
\end{equation}
It has been shown in \cite{6807812} that the following weighted Cheeger's inequalities hold:\begin{equation}
\lambda_2(\mathcal{L}_\mathbf{W})/2 \le h_\mathbf{W}(\mathcal{L}_\mathbf{W}) \le \sqrt{2 \delta_{\max} \lambda_2(\mathcal{L}_\mathbf{W})/w_{min}},
\end{equation}
where $\delta_{\max}$ is the maximum node degree, and $w_{\min}= \underset{i}\min ~w_i$.
By maximizing  $\lambda_2(\mathcal{L}_\mathbf{W})$,
 the lower bound of the weighted Cheeger constant increases. Hence, the ABS nodes can change their locations in order to maximize $\lambda_2({\mathcal{L}_\mathbf{W}})$ and thus $h_{\mathbf{W}}(\mathcal{L})$.

\subsection{ABSs Moving  Directions Towards Maximum Flow}\label{moveq}

\noindent In order to maximize the weighted  algebraic connectivity $\lambda_2({\mathcal{L}_\mathbf{W}})$, each ABS can move along the spatial gradient of $\lambda_2({\mathcal{L}_\mathbf{W}})$. Given the instantaneous position of an ABS node $i$, its spatial gradient along $x$-axis can be obtained~as follows:
\begin{align}
\frac{\partial \lambda_2({\mathcal{L}_\mathbf{W}}) }{\partial x_i} & =    {\mathbf{x}^f}^T \frac{\partial ({{\mathcal{L}_\mathbf{W}}}) }{\partial x_i} {\mathbf{x}^f}   \nonumber  \\
 &= \sum_{\{p,q:p \sim q \}}  \left[\frac{x_p^f}{\sqrt{w_p}}-\frac{x_q^f}{\sqrt{w_q}}\right]^2\frac{\partial a_{p,q}}{\partial x_i}, & 
\label{eq:moving}
\end{align}
where $\mathbf{x}^f$ is the Fiedler vector (i.e.,  the eigenvector corresponding to the second smallest eigenvalue), ${x}^f_p$ ($\mathbf{x}^f_q$) is the $p^\textrm{th}$ ($q^\textrm{th}$) component of $\mathbf{x}^f$, and $p \sim q$ means that
nodes $p$ and $q$ are connected. In order to compute \eqref{eq:moving}, we need to compute $\frac{\partial a_{p,q}}{\partial x_i}$, and it should be noted that if $p=q$, $\frac{\partial a_{p,q}}{\partial x_i}=0$.  Specifically:
\begin{align}
\frac{\partial a_{p,q}}{\partial x_i}&=B\left(\frac{1}{\ln(1+\textrm{SIR}_{p,q})+\ln(1+\textrm{SIR}_{q,p})}\right)^{-2} \nonumber  \\
  \begin{split}
  \hspace{1.5mm} \times \left[ \frac{1}{(1+\textrm{SIR}_{p,q})\ln^2(1+\textrm{SIR}_{p,q})}\frac{\partial \textrm{SIR}_{p,q}}{\partial x_i} \right. \\
    \left. + \frac{1}{(1+\textrm{SIR}_{q,p})\ln^2(1+\textrm{SIR}_{q,p})}\frac{\partial \textrm{SIR}_{q,p}}{\partial x_i} \vphantom{\int_1^2} \right].
    \end{split} & 
\label{eq:1}
\end{align}
For coordinates $y$ and $z$, the moving directions can be obtained similarly. 

	

\section{Multi-Commodity Maximum Flow Problem}\label{sec4}
\noindent In this section, the ABS interference-avoidance problem is investigated under both the multicast and the multi-unicast scenarios \cite{leighton1999multicommodity}. The maximum concurrent flow is defined as maximum of $f^m$    so that  $f^m D(k)$  units from commodity $1\le k \le {K}$  can be transferred to the destination  at the same time in which $D(k)$ is the demand for the commodity $k$. In the multi-commodity problem, typically there are more than one destination in the network \cite{leighton1999multicommodity}. In the multicast scenario, the network conducts the flow from the source to multiple destinations, while in the multi-unicast scenario, there are multiple pairs of source-destination with individual flows for each pair. The minimum multicut $C^m$ for the multi-commodity problem is defined as~{\cite{leighton1999multicommodity}:
\begin{equation}
C^m=\underset{S}{\text{min}} \frac {\sum_{i \in S, j \in \bar S}a_{i,j}}{\sum_{k\in \mathcal{K}(S)}D(k)},
\end{equation}
where $\mathcal{K}(S)=\{k:v_{s_k} \in S ~\textrm{and} ~ v_{d_k} \in \bar{S},~ \textrm{or}~ v_{s_k} \in \bar{S} ~ \textrm{and}~ v_{d_k} \in~S \}$, and $v_{s_k}$ and $v_{d_k}$ are the source and the destination for commodity $k$, respectively. Similar to the approach presented in Section~\ref{cheeg}, by defining $\mathbf{W}^{(k)}=\textrm{diag}\{w_1^{(k)}, ...,w_n^{(k)}\}$, the multi-weighted Cheeger  constant \cite{6807812}  can be defined as:
\begin{equation}
h^{(K)}_{\mathbf{W}}(\mathcal{L})=\underset{S}{\text{min}} \frac {\sum_{i \in S, j \in \bar S}a_{i,j}}{\sum_{k=1}^{K} \gamma_k\text{min}\{ |S|_{\mathbf{W}^{(k)}}),(|\bar S|_{\mathbf{W}^{(k)}})\}},
\end{equation}
where $\gamma_k = \min^{-p} \{ |S|_{\mathbf{W}^{(k)}}),(|\bar S|_{\mathbf{W}^{(k)}})\}$  for commodity $k = 1, ..., K$. Parameter $\gamma_k$  is inversely proportional to the $k^\textrm{th}$ flow bottleneck. The value of $p$ is related to the weight ${\mathbf{W}^{(k)}}$ and  its most suitable value can be obtained via numerical explorations. Defining $\bar {\mathbf{W}}=\textrm{diag}\{\bar w_1, ...,\bar w_n\}$ where $\bar w_i= \sum_{k=1}^{K}(w_i^{(k)})^{1-p}$, the weighted Laplacian matrix is given by $\mathcal{L}^{(K)}_{\bar{\mathbf{W}}}={\bar{\mathbf{W}}}^{-1/2}\mathcal{L} {\bar{\mathbf{W}}}^{-1/2}.$ As in Section~\ref{moveq}, the ABSs can take $\lambda_2 \left(\mathcal{L}^{(K)}_{\bar {\mathbf{W}}}\right)$  as a metric to reconfigure their positions,  so as to achieve a better performance as compared to the case of using $\lambda_2(\mathcal{L})$.

\section{Practical Considerations}\label{sec5}
\subsection{ABSs with Limited Communication Range }
\noindent The discussions in the previous sections assume that the network graph is a complete graph, which  may not be true in some scenarios due to the limited communication range $R_i^{\textrm{th}}$ of the ABSs.   The value of $R_i^{\textrm{th}}$ depends on the transmission power of ABS $i$. To capture this effect, it is assumed in this subsection that, for two given ABSs $i$ and $j$,  a communication link (or equivalently, an edge) exists only if $d_{i,j} \le R_i^{\textrm{th}}$. Hence, the definition of the adjacency matrix $\mathbf{A}$  of the ABS network graph is modified accordingly as
\begin{equation}\label{eq20}
a_{i,j}=\left\{\begin{matrix}
c_{i,j},& \textrm{if} ~i \ne j ~ \textrm{and} ~ d_{i,j} \le R_i^{th}\\ 
0, & \textrm{otherwise},
\end{matrix}\right.
\end{equation}
where $c_{i,j}$ is defined as in (1).


\subsection{Distributed Algorithm for Computing ABS  Directions}

\noindent As can be seen from \eqref{eq:moving},  to compute the moving directions, each ABS $i$ needs the Fiedler  vector $\mathbf{x}^f$ of the Laplacian matrix. To compute the Fiedler vector in a centralized way, each ABS needs to know the whole matrix ${\mathcal{L}_\mathbf{W}}$ at each time-slot which is  inefficient both in terms of communication overhead and computation, especially when the number of  ABSs is large. However, having a closer look at \eqref{eq:moving}, it can be seen that  ABS $i$ does not need all the elements of  $\mathbf{x}^f$ to compute its moving direction. It just needs the elements of $\mathbf{x}^f$  corresponding to its neighbors. As a result, a distributed algorithm can be used to compute the Fiedler vector to reduce both the  communication  and the computation complexity for the ABS nodes.

Based on the algorithm proposed in \cite{bertrand2013distributed}, the Fiedler vector  can be
computed  in a distributed fashion. During each iteration, the nodes are assumed
to perform a pre-defined task and share the result with their
neighbors. When the algorithm is done, 
each node eventually has access to its corresponding
entry in the Fiedler vector. By increasing the number of iterations, the error of the computed Fiedler vector is decreased. It is worth mentioning that to compute the moving direction for the ABS $i$, the elements of the Fiedler vector of its neighbors are needed. Here, we  assume that the ABSs can exchange such information with their neighbors. Algorithm~\ref{alg} describes the details of the proposed distributed algorithm based on \cite{bertrand2013distributed}.

\subsection{Energy Consumption and Maximum Flow Trade-off}\label{sec6}
\noindent Since the ABSs are using batteries as their energy sources, developing energy efficient trajectory is critical. However, the optimal trajectories for maximum flow of the network are not necessarily the best trajectories for energy saving. If the ABSs compute their moving directions solely using the maximum flow criterion in each iteration,   the trajectory between the initial and the final points is usually not the most energy-efficient. In some applications, the  ABSs can move in a more energy-efficient way at the expense of a longer operation time.

 \begin{algorithm}[t]
	
	\DontPrintSemicolon
	\textbf{Initialization:} $B$, $r_{int}$, $\mathcal{T}$.
    
    For each ABS $i \in \mathcal{N}$, the distance $d_{i,j}$ to its neighbor $j \in \mathcal{N}_i$ is given. If the ABSs are neighbors, they can transfer this information easily via direct link. If the ABS nodes are not neighbors, the information can be transferred via a shortest path between the nodes.

	 \For{ $t = 1, 2, ..., \mathcal{T}$ }{       
        {

    Each ABS $i$ computes its corresponding entry in $\mathbf{x}^f$ using the distributed algorithm in \cite{bertrand2013distributed}.
    
    Each ABS shares the corresponding entry in $\mathbf{x}^f$  with its neighbors $\mathcal{N}_i$.
    
    Each ABS $i$ needs to compute $\frac{\partial a_{p,q}}{\partial x_i}$ for its neighbors as in \eqref{eq:1}.
    
    Each ABS $i$ computes its moving directions  based on \eqref{eq:moving}.
    
    Each ABS $i$ updates its location based on the obtained moving directions till convergence.

     } }

	\caption{{Distributed Algorithm for Moving Direction Computation for the ABSs}}
	\label{alg}
\end{algorithm}

To achieve the desired energy-efficient mobility management for the ABSs, suppose that each time-slot $t$ consists of $t^\textrm{c}$ (the \textit{computation time}) and $t^\textrm{m}$ (the \textit{moving time}). The computation time is the amount of time needed to compute the moving directions as in \eqref{eq:moving} which, for ABS $i$,  can be obtained as $t^\textrm{c}_i=\frac{K_i^2}{e_i}$ \cite{mozaffari2018beyond}; here, $K_i$ is the amount of data that must be processed before movement and $e_i$ is the processing speed of  ABS $i$. The moving time is the amount of time needed for each ABS to traverse the trajectory which is equal to $t_i^\textrm{m}=\frac{D_i}{v_i}$  \cite{di2015energy}; here, $D_i$ is the length of the path traversed by ABS $i$ and $v_i$ is the ABS speed.  Let us assume that  the ABSs try to compute their final locations in  $L$ iterations (time slots)  according to the maximum flow metric. Having the final location using maximum-flow criterion, the ABSs can employ an energy efficient trajectory to reach their final locations. This imposes $Lt_i^\textrm{c}$ latency to the network before they start to move as compared to the maximum flow trajectory.

More specifically, denote the initial and the final locations of the ABS $i$ obtained using the method in Section IV-C by $q_i[0]$, and $q_i[L]$, respectively. Here, we assume the ABSs can converge to the network maximum flow in $L$ iterations  using the weighted Cheeger constant approach. The ABSs can wait $Lt_i^\textrm{c}$ amount of time to compute  $q_i[L]$ and then choose the most energy efficient trajectory to traverse to the final location. As a result, the energy efficient approach imposes $Lt_i^\textrm{c}$ amount of computation delay to the network before the ABSs start to move. After reaching the final location, ABSs can relay the information with the maximum flow. 

\begin{figure*}[ht!]
   \subfloat[\label{fig1a}]{%
      \includegraphics[ width=5.65cm,height=6.3cm]{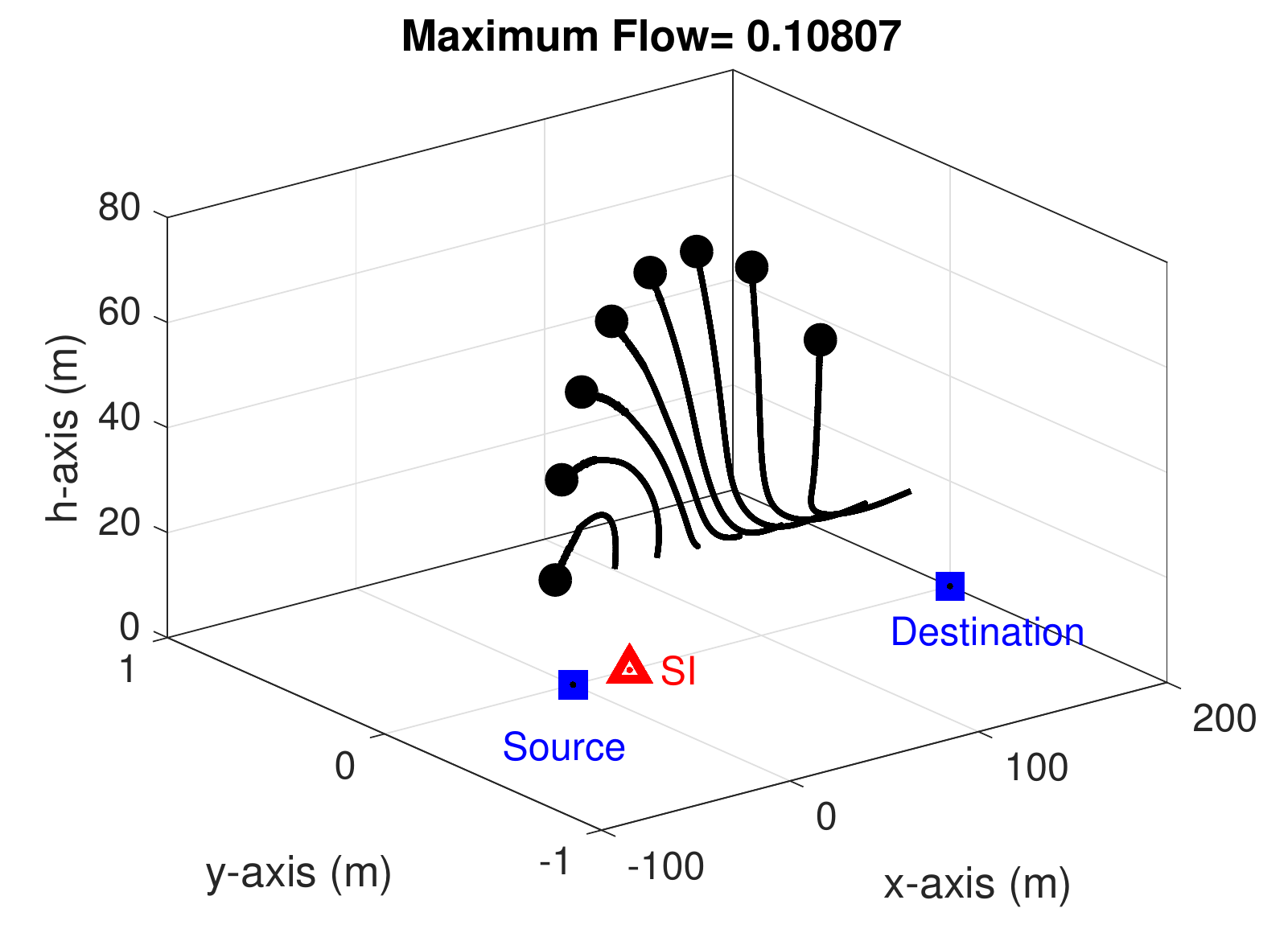}}
\hspace{\fill}
   \subfloat[\label{fig1b} ]{%
      \includegraphics[width=5.85cm,height=6.3cm]{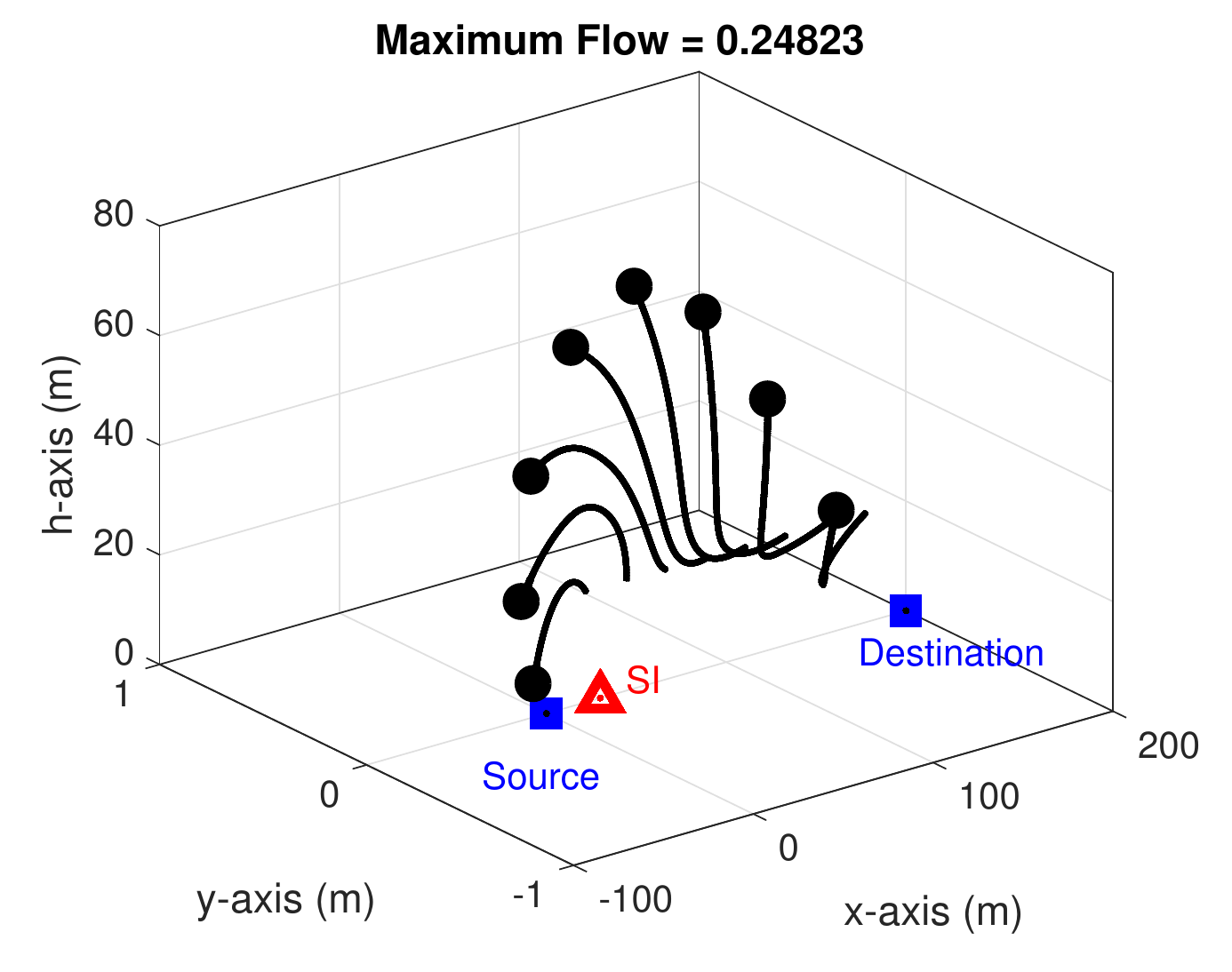}}
\hspace{\fill}
   \subfloat[\label{fig1c}]{%
      \includegraphics[width=5.95cm,height=6.2cm]{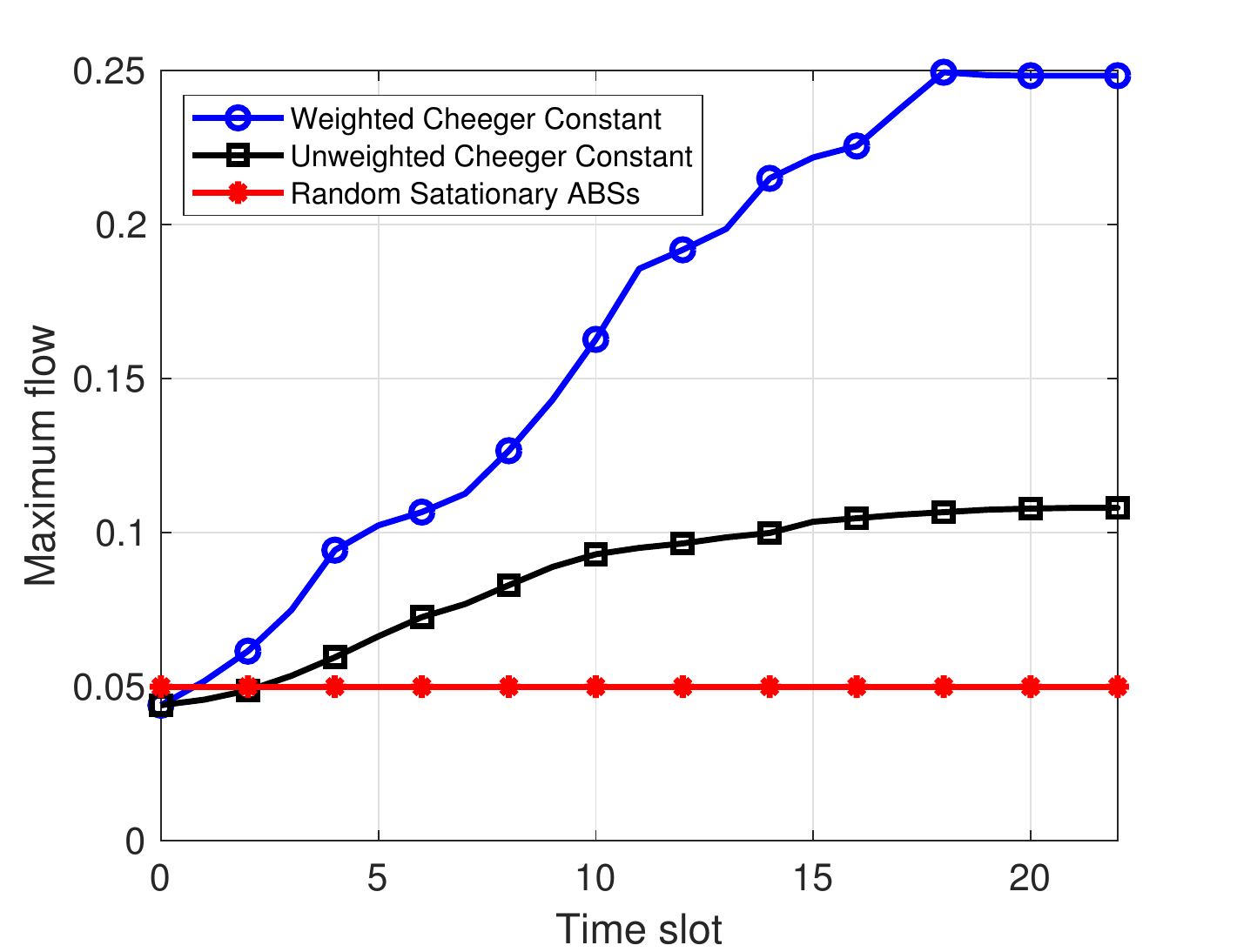}}\\
\caption{\label{costrr} The trajectories of the ABSs in the presence of 1 SI based on: (a)  the unweighted Cheeger constant and (b) the weighted Cheeger constant. (c) Comparison of the maximum flows throughout the network for movement of the ABSs based on the weighted/unweighted Cheeger constant and random stationary ABSs.}
\end{figure*}
 Assuming rotary wing ABSs, the energy consumption of each ABS $i$  between two stopping positions with  path length $D_i$ can be computed as~\cite{di2015energy}:
\begin{equation}\label{eq19}
E_i=\frac{D_i}{v_i}\left(P^\textrm{V}_i+P^\textrm{H}_i \right),
\end{equation}
where  $v_i$ is the ABS speed, $P^\textrm{V}_i$ and $P^\textrm{H}_i$ are the power consumption for vertical and horizontal movement, respectively. Denoting the height difference between two stopping locations by $\Delta h_i$, the vertical and the horizontal velocities can be computed as $v^v_i=v_i \sin \phi_i$ and $v_i^h=v \sin \phi_i$, where $\phi_i=\sin^{-1} \left(\frac{\Delta h_i}{D_i}\right)$. The horizontal power consumption $P^H_i$ is the summation of the parasitic power and the induced power for overcoming the parasitic drag. The parasitic power is given by~\cite{7888557}:
\begin{equation}\label{eq20}\nonumber
P^P_i=\frac{1}{2}\rho C_{D_o}A_e (v^h_i)^3+\frac{\pi}{4} N_b c_b \rho C_{D_o}\omega_i^3 R^4 \left(1+3 \left(\frac{v^h_i}{\omega_i R}\right)^2\right),
\end{equation}
where $C_{D_o}$ is the drag coefficient, $\rho$ is the air density, $c_b$ is the blade chord, $N_b$ is the number of blades, and $A_e$ is the reference area. The induced power can be obtained by:
\begin{equation}
P^I_i=\omega_i R W \mu_i,
\end{equation}
where $R$ is the motor disk radius, $W$ is the ABS weight, and $\omega_i$ is the blade angular velocity, and $\mu_i$ can be obtained using:
\begin{equation}
g(\mu_i)=2\ \rho \pi \omega^2 R^4 \mu_i \sqrt{\frac{(v^h_i)^2}{\omega_i^2 R^2}+ \mu_i^2}-W.
\end{equation}
The power consumption for vertical movement in the climbing and the descending cases is given by {\cite{mozaffari2017mobile}}:
\begin{equation}
P^V_i=\left\{\begin{matrix}
\frac{W}{2}v^v_i + \frac{W}{2}\sqrt{(v^v_i)^2+ \frac{2 W}{\rho \pi R^2}},& \textrm{climbing} \\ 
\frac{W}{2}v^v_i - \frac{W}{2}\sqrt{(v^v_i)^2+ \frac{2 W}{\rho \pi R^2}},& \textrm{descending}.
\end{matrix}\right.
\end{equation}
As can be seen from \eqref{eq19}, the energy consumption of the ABSs is proportional to the moving time ($\frac{D_i}{v_i}$), $P^V_i$ and $P^H_i$. Since the ABSs should reach to their final locations at the same time, we assume equal moving times for the ABSs at each time-slot. Base on this, the energy consumption depends on the  summation of $P^V_i$ and $P^H_i$ which  is a non-decreasing function of the ABSs' speed.  As a result, given the initial and the final locations after computing the moving directions as in \eqref{eq:moving}, the most energy efficient path is the straight line connecting the  two locations $q_i[0]$ and $q_i[L]$; the reason is that, considering the  fixed moving time in each time-slot,  the minimum distance ABS trajectory can result in the minimum speed for ABSs.
\begin {table}[t!]
\caption {Simulation parameters.} \label{tab1} 
\vspace{-0.4cm}
\begin{center}
\begin{tabular}{ |c|c|c|c| }
\hline
\textbf{Parameter} & \textbf{Description} & \textbf{Value} \\
\hline
\footnotesize{$f_c$} & \footnotesize{Carrier frequency} & \footnotesize{2 GHz}  \\ \hline
\footnotesize{$B$} & \footnotesize{Bandwidth} & \footnotesize{1 Hz} \\ \hline \footnotesize{$v_{\textrm{max}}$} & \footnotesize{ABS maximum speed} & \footnotesize{5 m/s} \\ \hline \footnotesize{$\zeta, \kappa$} & \footnotesize{Smoothed step function parameters} & \footnotesize{1, 10} \\ \hline \footnotesize{$r_{\textrm{int}}$}
& \footnotesize{Interference radius} & \footnotesize{5 m} \\ \hline \footnotesize{$\mu_{\textrm{LoS}}$,
$\mu_{\textrm{{NLoS}}}$
} & \footnotesize{Attenuation factors for LoS/NLoS } & \footnotesize{5 dB, 20 dB} \\ \hline \footnotesize{$\psi$, $\eta$} & \footnotesize{Constant values for $P^\textrm{LoS}_{i,j}$} & \footnotesize{11.95, 0.14} \\ \hline \footnotesize{$\rho$} & \footnotesize{Air density} & \footnotesize{1.225 kg/$\textrm{m}^{-3}$} \\ \hline
\footnotesize{$\omega$} & \footnotesize{Angular velocity} & \footnotesize{20 rad/sec} \\ \hline
\footnotesize{$R$} & \footnotesize{Rotor disk radius} & \footnotesize{0.5 m} \\ \hline
\footnotesize{$c_b$} & \footnotesize{Blade chord} & \footnotesize{10 cm} \\ \hline
\footnotesize{$N_b$} & \footnotesize{Number of blades} & \footnotesize{4} \\ \hline
\footnotesize{$W$} & \footnotesize{Weight of ABS} & \footnotesize{50 N} \\ \hline

\end{tabular}
\end{center}
\end {table}
\begin{figure*}[ht!]
   \subfloat[\label{fig4a}]{%
      \includegraphics[ width=5.95cm,height=6.4cm]{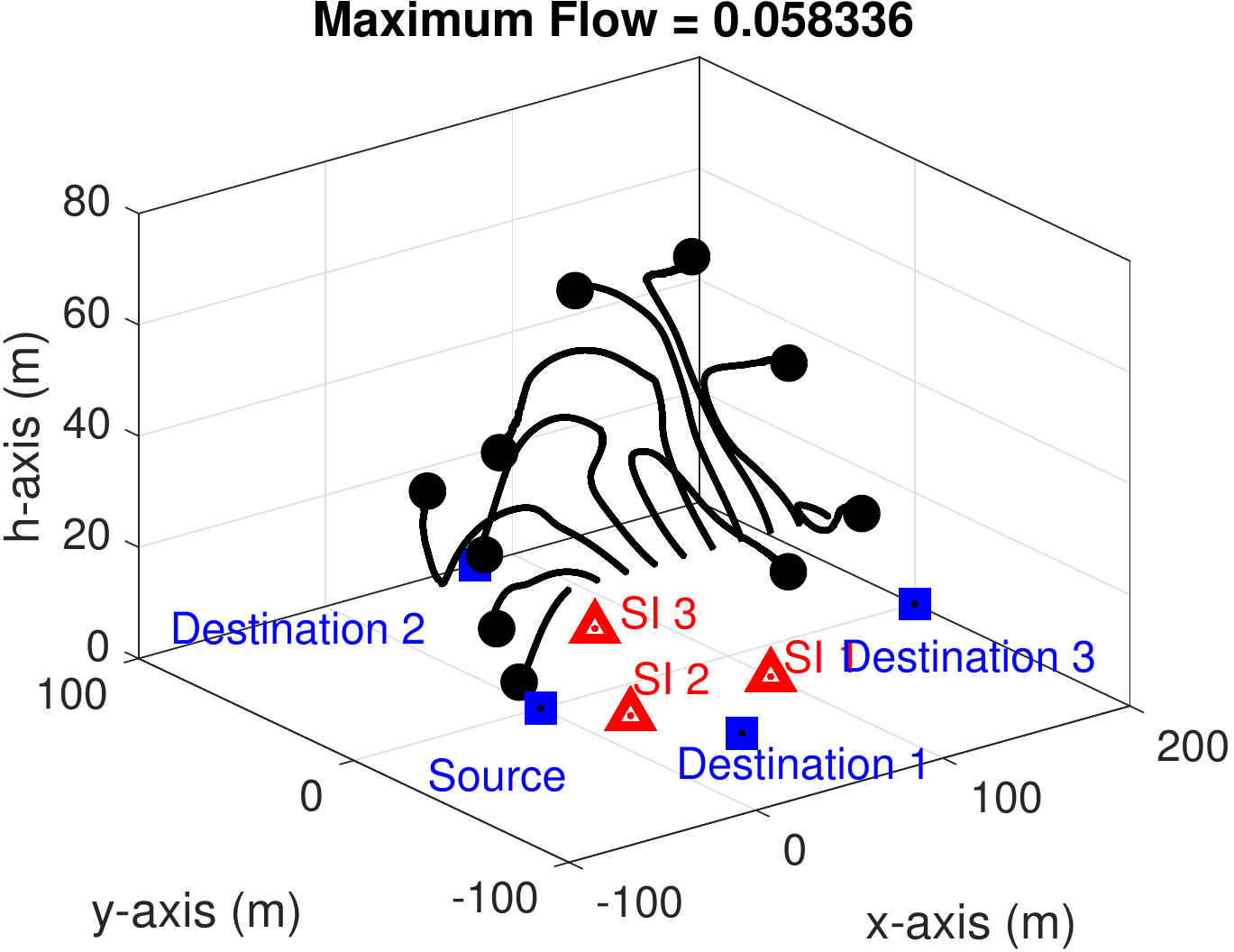}}
\hspace{\fill}
   \subfloat[\label{fig4b} ]{%
      \includegraphics[width=5.95cm,height=6.3cm]{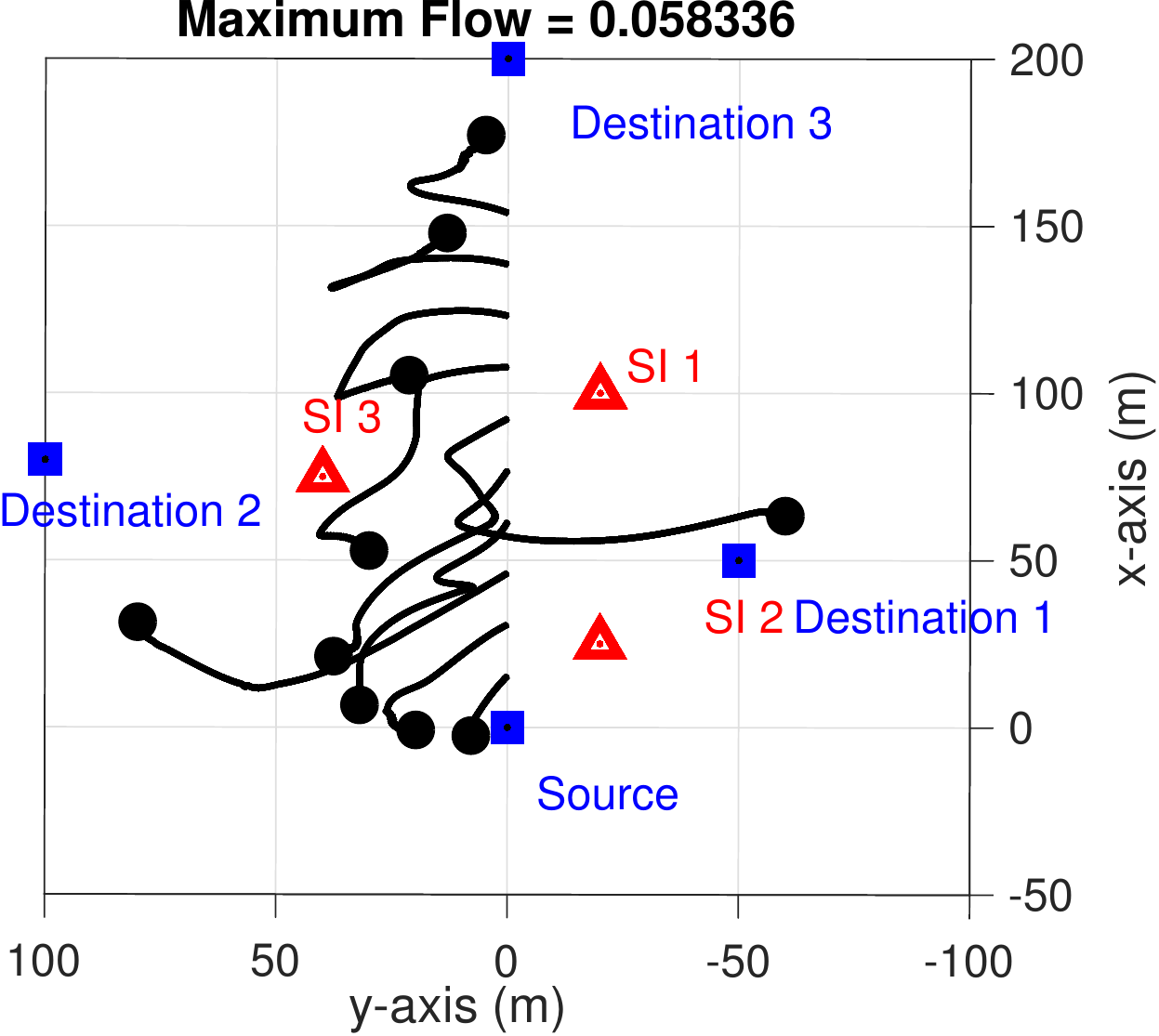}}
\hspace{\fill}
   \subfloat[\label{fig4c}]{%
      \includegraphics[width=5.95cm,height=6.4cm]{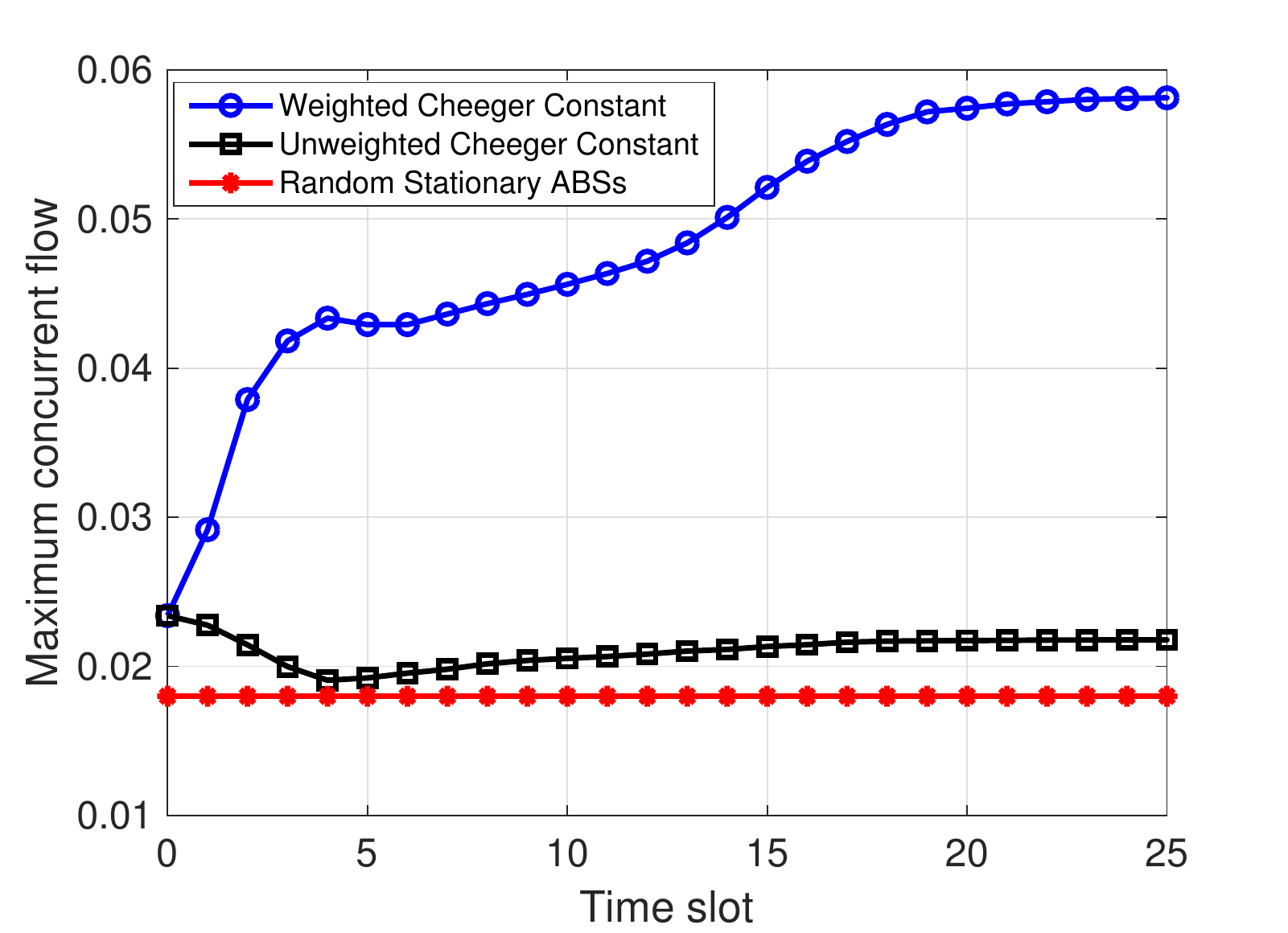}}\\
\caption{\label{fig4} The trajectories of the ABSs  in the presence of 3 SIs using the weighted Cheeger constant for a multi-cast scenario: (a)  3D view, (b) Top view. (c) Comparison of the maximum flow throughout the network for movement of the ABSs based on the weighted/unweighted Cheeger constant and random stationary ABSs.}
\end{figure*}

\begin{figure*}[ht!]
   \subfloat[\label{fig5a}]{%
      \includegraphics[ width=5.95cm,height=6.4cm]{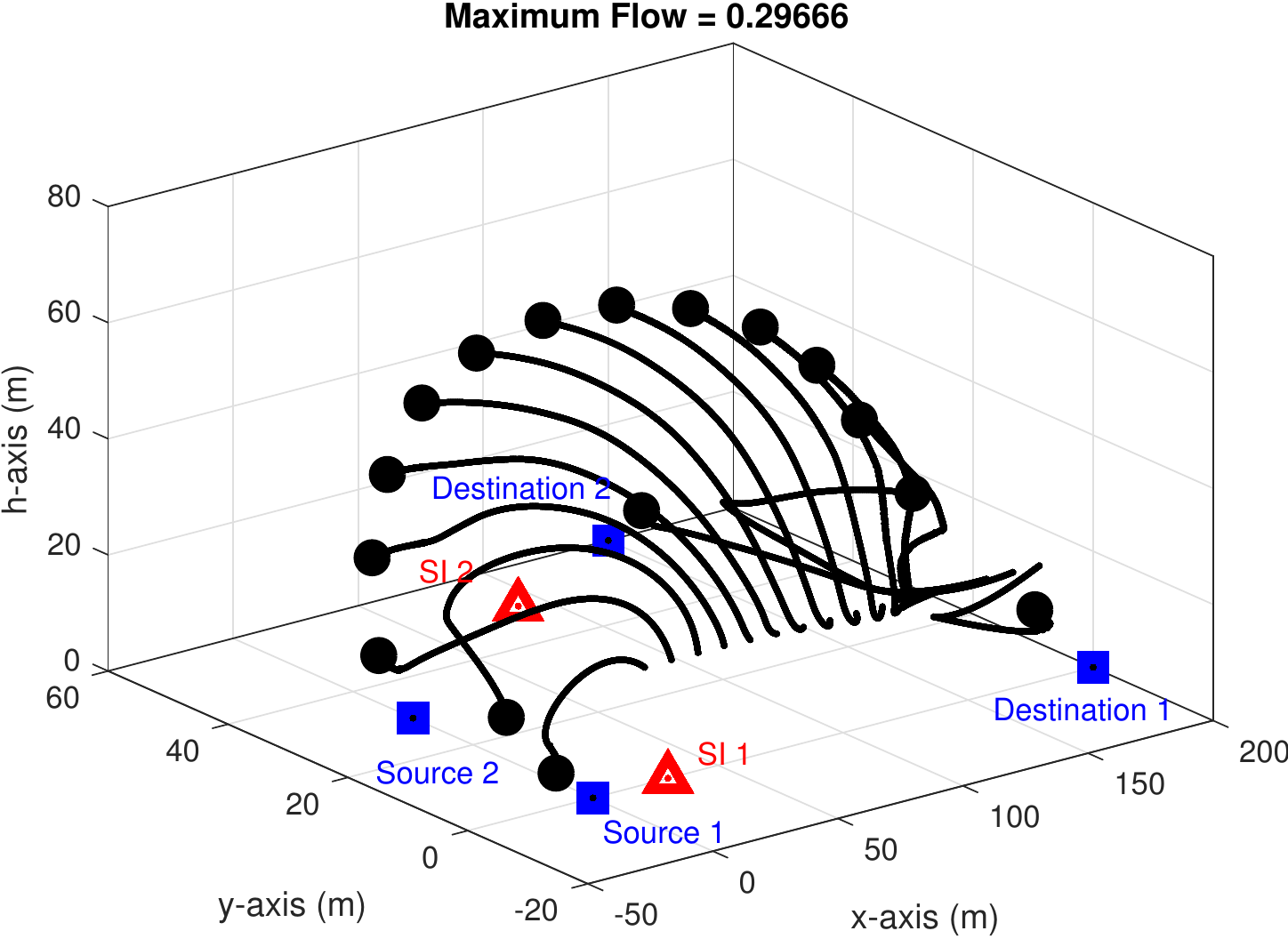}}
\hspace{\fill}
   \subfloat[\label{fig5b} ]{%
      \includegraphics[width=5.75cm,height=6.2cm]{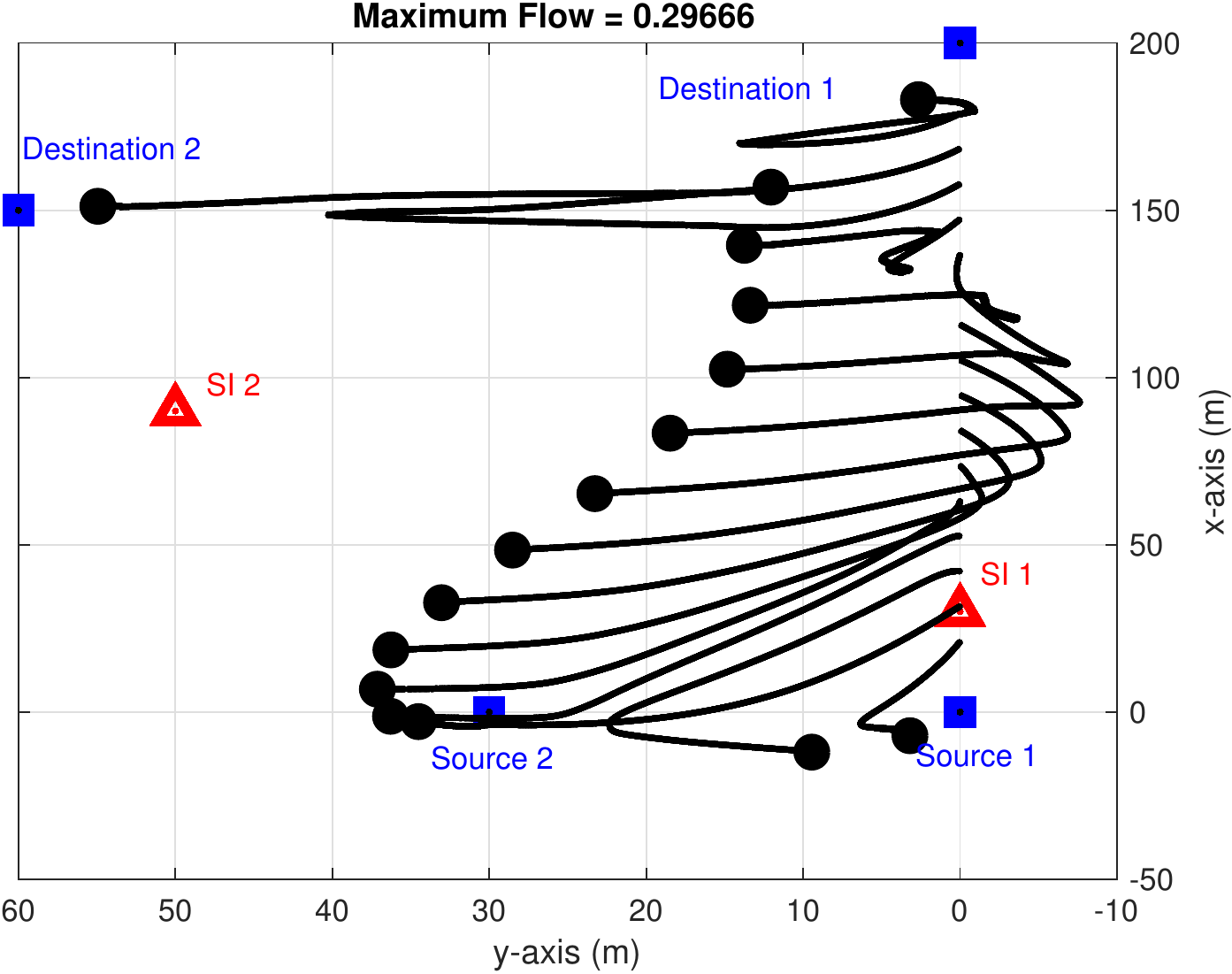}}
\hspace{\fill}
   \subfloat[\label{fig5c}]{%
      \includegraphics[width=5.95cm,height=6.4cm]{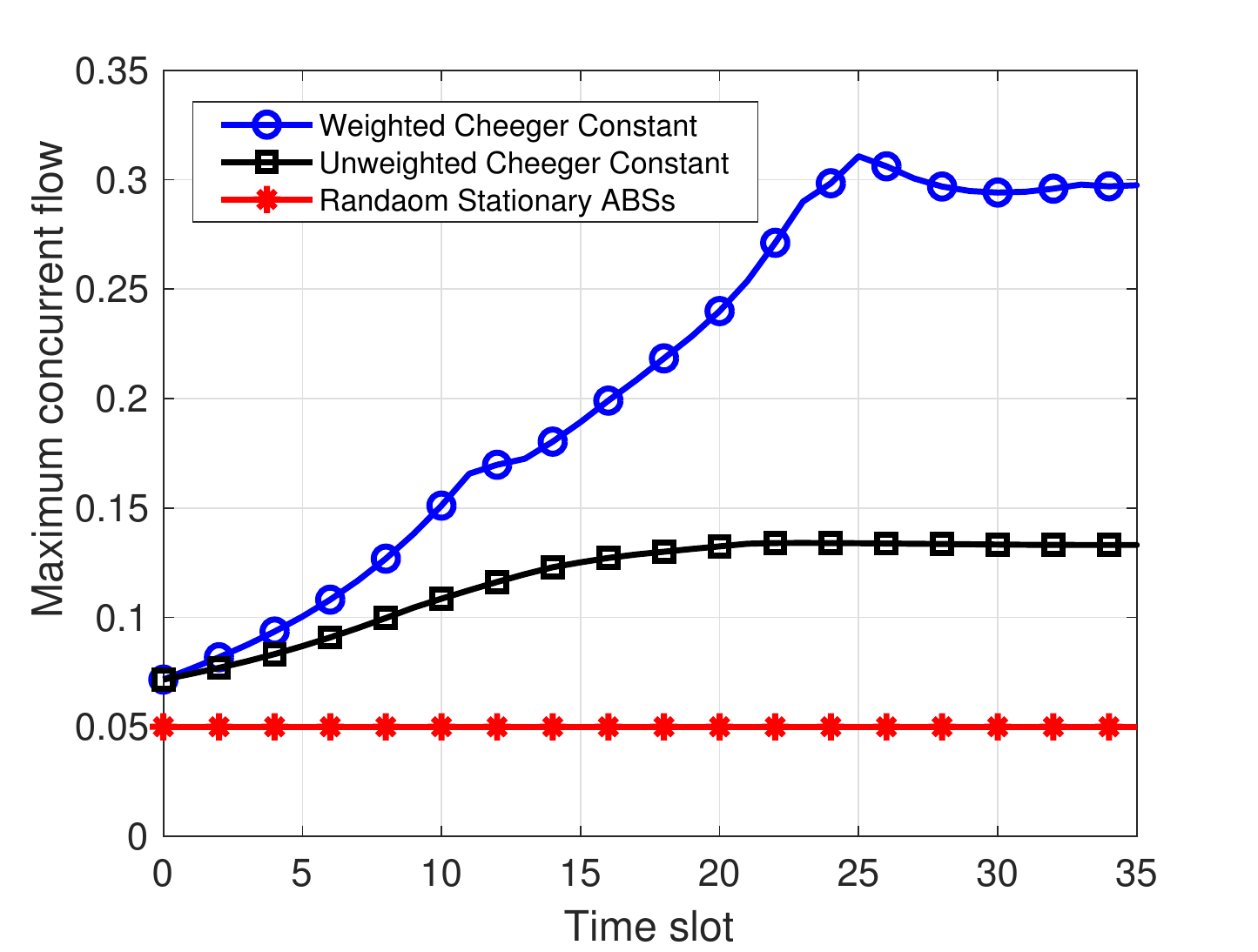}}\\
\caption{\label{fig5} The trajectories of the ABSs  in the presence of 2 SIs using the weighted Cheeger constant for a multi-unicast scenario: (a)  3D view, (b) Top view. (c) Comparison of the maximum flow throughout the network for movement of the ABSs based on the weighted/unweighted Cheeger constant and random stationary ABSs.}
\end{figure*}
\section{Simulation 
Results}\label{sec7}

\noindent We have conducted extensive simulations  to evaluate the effectiveness of the proposed dynamic  mobility-aware interference avoidance scheme. The parameters are presented in Table \ref{tab1}, unless otherwise stated. In the following, we consider both single-commodity and multi-commodity maximum flow scenarios.

In the single-commodity case, a pair of source and destination aim to communicate through several ABSs in the presence of SIs from the primary network.
Consider a 3D cubic space of size 200 m $\times$ 200 m $\times$ 100 m, in which 8 ABSs are deployed. Each ABS $i \in \{1, 2, \dots, 8\}$ is initially located at $(0,25i,20)$ m. That is, the ABSs are in a straight line at the height of $20$ meters from the ground with equal distances of 25 m in the $x$ coordinate.  The source and the destination are fixed on the ground at positions $(0,0,0)$ m and $(200,0,0)$ m.
The moving directions of the ABSs are computed using weighted and unweighted Cheeger constants. We use random stationary ABS positioning  as a baseline for comparison. In this baseline, the ABSs are randomly positioned on a 200 m $\times$ 200 m plane with a fixed height of 20 m. All the simulation results are averaged over 1000 Monte Carlo runs.

As seen in Fig.~1(a)  and Fig.~1(b), one SI is located at  $(30,0,0)$ m. The  ABSs are moving in a 3D  space. The black circles and black lines  represent the ABSs and their trajectories towards maximum flow locations, respectively.  As seen in Fig.~1(c), the  mobility of the ABSs can considerably increase the maximum flow as compared to stationary deployment. Moreover, the weighted Cheeger constant  further improves the maximum flow by about $150 \%$ as compared to the conventional Cheeger constant.


In Fig. \ref{fig4}, a multi-cast scenario is considered in which one source aims at transmitting  data to three destinations in the presence of 3 SIs. Here, 12 ABSs are considered. Fig.~2(a) shows the 3D view of the ABS trajectories, while the top view in Fig.~2(b) can help better illustrate the final locations. Fig.~2(c) shows the effectiveness of the weighted Cheeger constant in comparison to the other two baselines.
\begin{figure}[!t]
	\includegraphics[width=8.5cm,height=7.6cm]{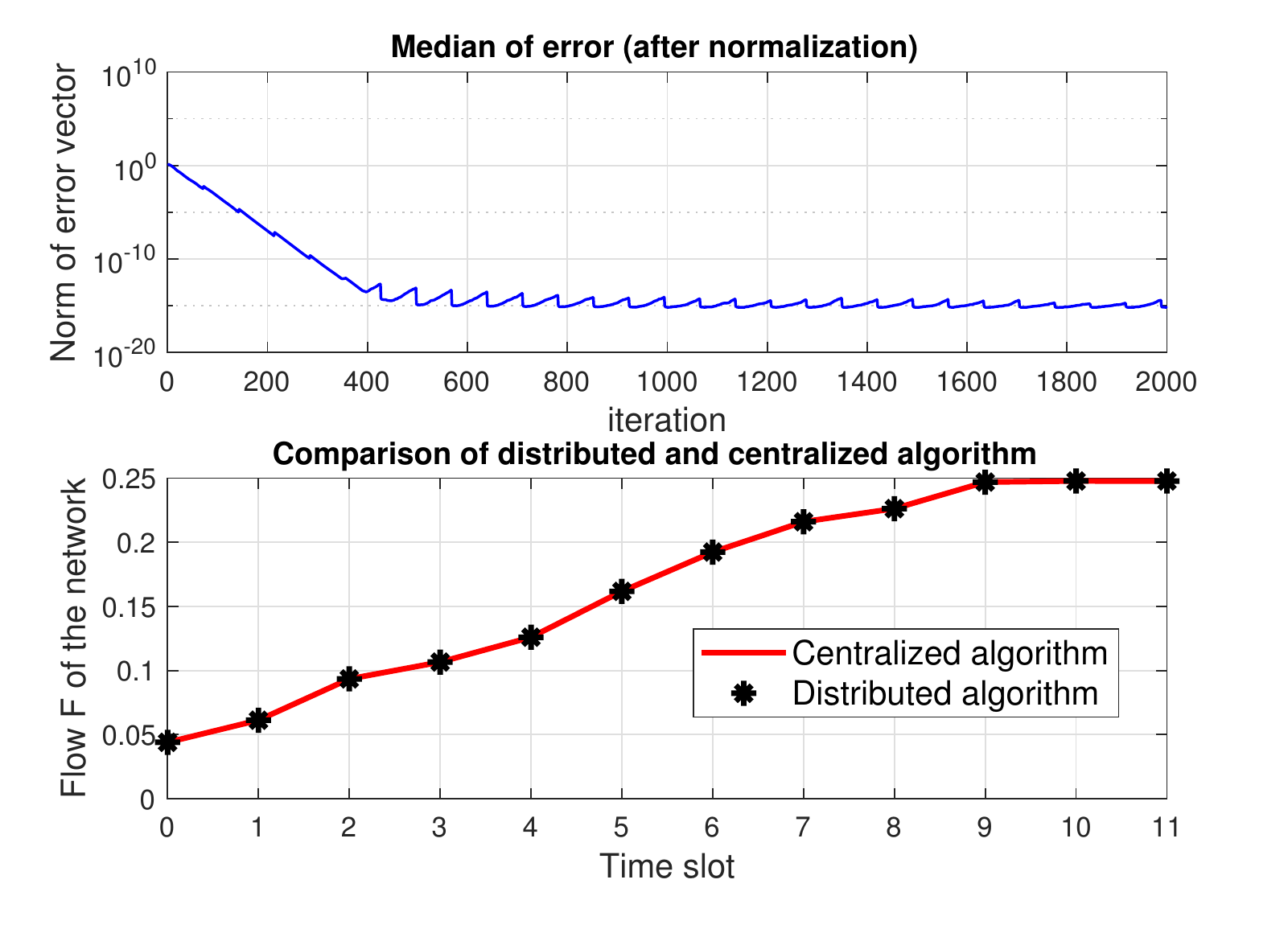}
	\centering
	\caption{Comparison of the distributed algorithm and the centralized algorithm for the computation of the Fiedler vector on the network maximum flow while deploying the weighted Cheeger constant.}
	\label{fig7}
\end{figure}
\begin{figure}[!t]
	\includegraphics[width=8.5cm,height=7.2cm]{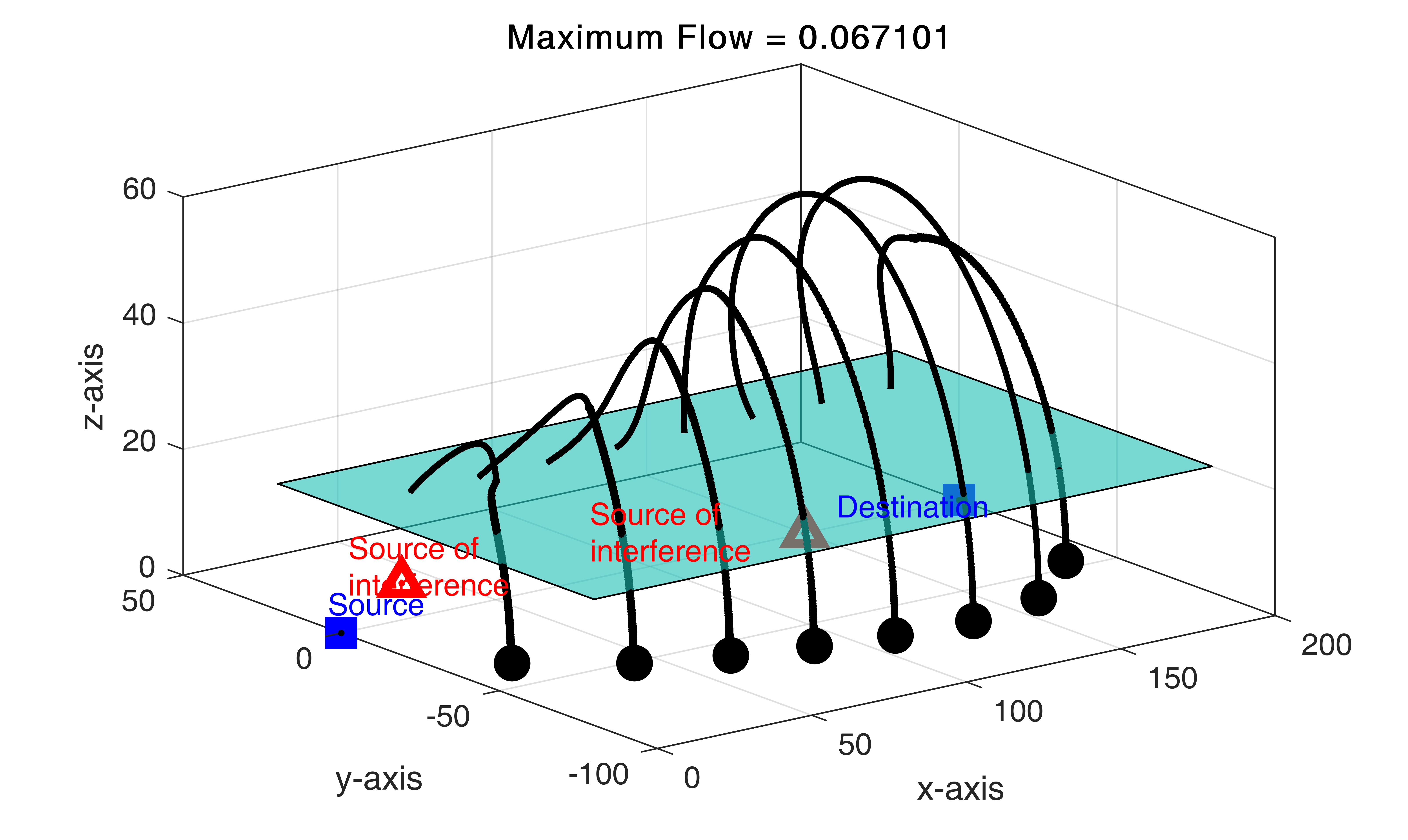}
	\centering
	\caption{Depending on the SI positions, in some scenarios, the ABSs move towards the positions with zero height.}
	\label{fig3}
\end{figure}
\begin{figure}[!t]
	\includegraphics[width=9cm,height=7.6cm]{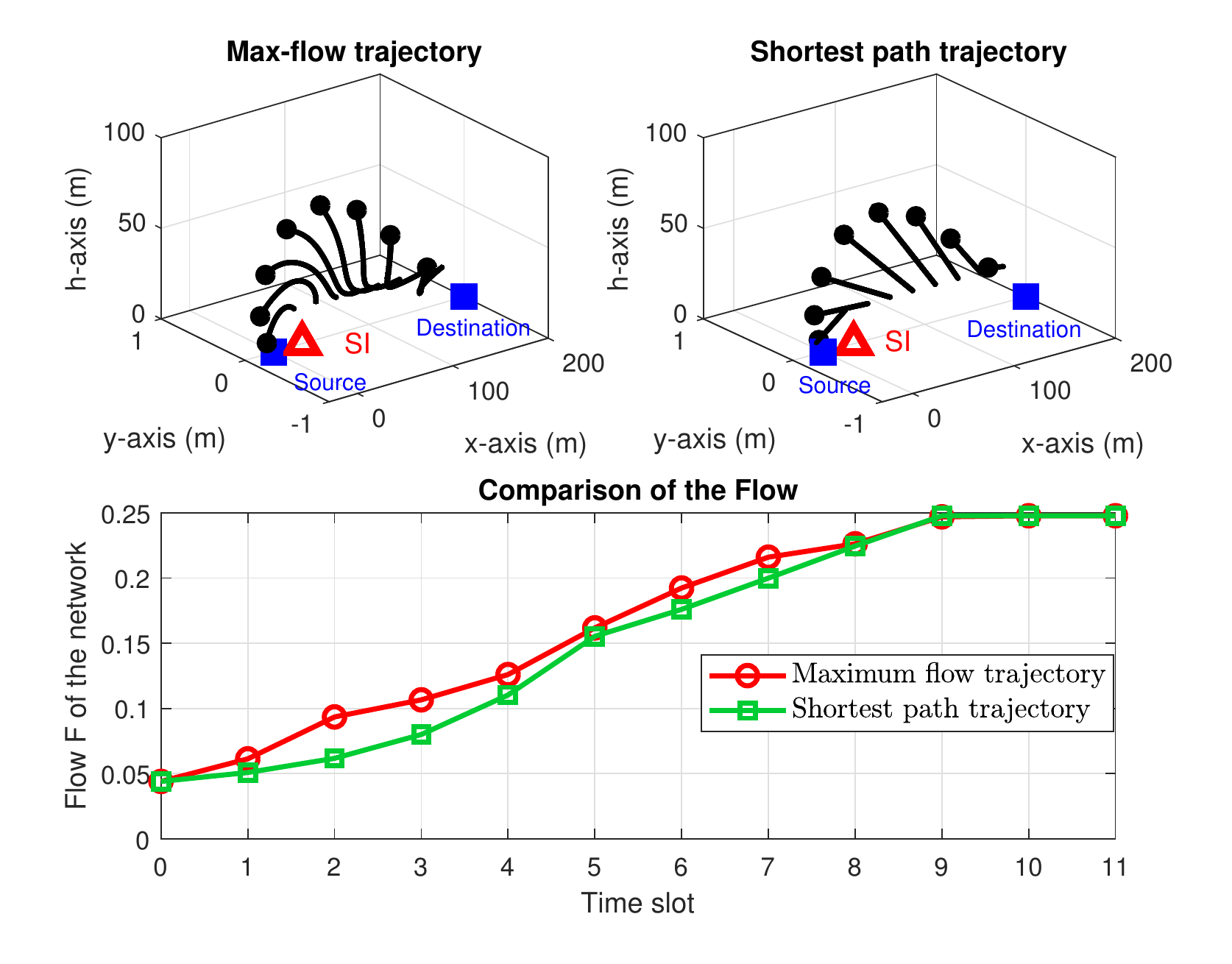}
	\centering
	\caption{Comparison of the network flows for maximum flow trajectory versus shortest path trajectory in a single-commodity scenario.}
	\label{eeflow}
\end{figure}

\begin{figure}[!t]
	\includegraphics[width=8cm,height=7.2cm]{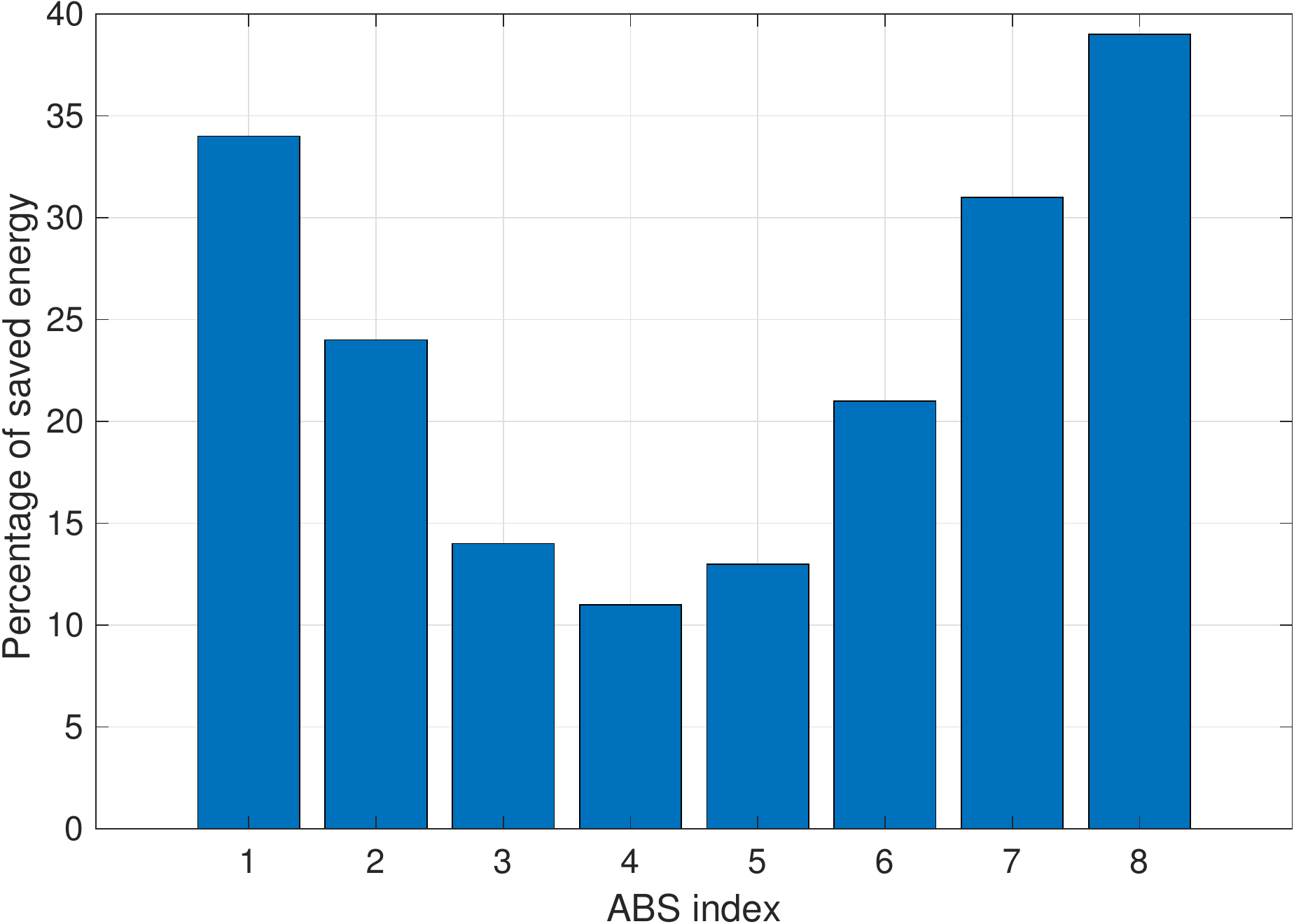}
	\centering
	\caption{Percentage of the saved energy  for different trajectories schemes for each ABS.}
	\label{tab2}
\end{figure}
In Fig. \ref{fig5}, a multi-unicast case is considered in which two pairs of  source and destination aim to communicate. Here, 16 ABSs are deployed. The 3D view in Fig.~3(a) and the top view in Fig.~3(b) present the positions of the ABSs towards the maximum flow. In Fig.~3(b), it can be seen that one of the ABSs is changing its direction toward destination 2 to form a better link for the data flow. As shown in Fig.~3(c), the weighted Cheeger constant outperforms the other two approaches.

Considering the single-commodity case scenario, in Fig.~\ref{fig7}, the maximum flows of the network are compared for the cases of centralized and distributed computation of the Feidler vector. In the top sub-figure, the norm of the error vector for the  Fiedler vector is computed over the iterations for the distributed algorithm.  As  can be seen, if the number of iterations is sufficiently large, the error of the distributed algorithm is negligible. Hence, the results obtained form  the distributed algorithm and the centralized algorithm are identical, as shown in the bottom sub-figure.

In simulations, if no constraints are imposed for the heights of the ABSs, it can be seen that, in some situations, the final heights of the ABSs are equal to zero, which is impractical. Such a case may occur when there is no SI on one side of the region of interest. To address this issue, we may impose a height constraint for each ABS as the green plane shown in Fig.~\ref{fig3} to prevent   the ABSs from moving to an amplitude lower than $h=20$ m. In the simulated scenario, this leads to a loss of 0.023 of maximum flow for the given source-destination pair.

In Fig.~\ref{eeflow}, the flows of the network with  the maximum flow trajectory (obtained by using the methods in Section~\ref{cheeg}) and the energy-efficient trajectory (obtained by using the method in Section~\ref{sec6}) are compared. The percentages of saved energy for different ABSs are compared in Fig.~\ref{tab2}, which is defined  as $\epsilon_i=100 \times (1-{E_i^{\textrm{Energy-efficient}}}/{E_i^{\textrm{Maximum-flow}}})$.  For this simulation setting, the single-commodity maximum flow problem is considered in the presence of one SI. The 3D plots for different trajectories are included  in Fig.~\ref{eeflow}. 
As can be seen from Fig.~\ref{eeflow}, the loss in flow of the network is temporary and insignificant. On the other hand, as shown in Fig.~\ref{tab2}, each ABS can save energy from about $11 \%$ to $39 \%$, depending on its initial location. One important observation is that for the ABSs that are in the middle during the initial deployment, the ratio of the saved energy is lower since their trajectories are closer to the shortest path.


\section{Conclusion}\label{sec8}
\noindent In this paper, an adaptive mobility aware interference avoidance scheme is considered  for ABSs in which the ABSs aim at forming a temporary network to send  information from a terrestrial source to a terrestrial destination in coexistence with a primary network. The primary network causes interference to the ABS temporary network. Due to the inherent feature of the ABSs, i.e., their mobility, they can reconfigure their locations to avoid the interference. Using the weighted Cheeger constant as the connectivity metric, the optimal moving directions to achieve the maximum flow is obtained for the ABSs. Furthermore, a distributed algorithm for computing  the moving directions is proposed, and the trade-off between maximum flow and energy efficient trajectories is investigated. The simulation results confirm the effectiveness of the proposed scheme, in which each ABS can save up to $39 \%$ of the energy  with marginal degradation in maximum flow.

\bibliographystyle{ieeetr}
\bibliography{refs}

\end{document}